\documentclass[aps,prl,showpacs,twocolumn,superscriptaddress,groupedaddress,amsmath,amssymb]{revtex4-1}
\usepackage[upright]{fourier}
\usepackage{upgreek}
\usepackage{latexsym}
\usepackage{amssymb}
\usepackage{lineno}
\usepackage{graphicx}
\usepackage{dcolumn}
\usepackage{multirow}
\usepackage{slashbox}
\usepackage{bm}
\usepackage{color}
\usepackage{subfigure}
\usepackage{xspace}
\usepackage{textpos}
\usepackage[english]{babel}

\newcommand{\jpsi}{J/\psi}
\newcommand{\pip}{\pi^+}
\newcommand{\pin}{\pi^-}

\newcommand{\ar}{\rightarrow}
\newcommand{\ra}{\rightarrow}
\newcommand{\pio}{\pi^0}

\renewcommand{\phi}{\upphi}

\begin{document}
\title{\boldmath Search for a strangeonium-like structure $Z_s$
decaying into $\phi \pi$ and a measurement of the cross section
$e^+e^-\rightarrow\phi\pi\pi$}
\author{
 \begin{small}
 \begin{center}
M.~Ablikim$^{1}$, M.~N.~Achasov$^{9,d}$, S. ~Ahmed$^{14}$, M.~Albrecht$^{4}$, A.~Amoroso$^{53A,53C}$, F.~F.~An$^{1}$, Q.~An$^{50,40}$, J.~Z.~Bai$^{1}$, Y.~Bai$^{39}$, O.~Bakina$^{24}$
% ...
\\
\vspace{0.2cm}
(BESIII Collaboration)\\
\vspace{0.2cm} {\it
$^{1}$ Institute of High Energy Physics, Beijing 100049, People's Republic of China\\
% ...
$^{4}$ Bochum Ruhr-University, D-44780 Bochum, Germany\\
% ...
$^{9}$ G.I. Budker Institute of Nuclear Physics SB RAS (BINP), Novosibirsk 630090, Russia\\
% ...
$^{14}$ Helmholtz Institute Mainz, Johann-Joachim-Becher-Weg 45, D-55099 Mainz, Germany\\
% ...
$^{24}$ Joint Institute for Nuclear Research, 141980 Dubna, Moscow region, Russia\\
% ...
$^{39}$ Southeast University, Nanjing 211100, People's Republic of China\\
$^{40}$ State Key Laboratory of Particle Detection and Electronics, Beijing 100049, Hefei 230026, People's Republic of China\\
% ...
$^{50}$ University of Science and Technology of China, Hefei 230026, People's Republic of China\\
% ...
$^{53}$ (A)University of Turin, I-10125, Turin, Italy; (B)University of Eastern Piedmont, I-15121, Alessandria, Italy; (C)INFN, I-10125, Turin, Italy\\
% ...
$^{d}$ Also at the Novosibirsk State University, Novosibirsk, 630090, Russia\\
% ...
}
\end{center}
\vspace{0.4cm}
\end{small}
}
\noaffiliation{}

\begin{abstract}
Using a data sample of $e^+e^-$ collision data corresponding to an
integrated luminosity of 108 pb$^{-1}$ collected with the BESIII
detector at a center-of-mass energy of 2.125 GeV, we study the
process $e^+e^-\rightarrow \phi\pi\pi$ and search for a
strangeoniumlike structure $Z_s$ decaying into $\phi\pi$. No
signal is observed in the $\phi\pi$ mass spectrum.
Upper limits on the cross sections for
$Z_s$ production at the 90\% confidence level are determined. In
addition, the cross sections of
$e^+e^-\rightarrow\phi\pi^{+}\pi^{-}$ and $e^+e^-\rightarrow
\phi\pi^{0}\pi^{0}$ at 2.125 GeV are measured to be
$(436.2\pm6.4\pm30.1)$ pb and $(237.0\pm8.6\pm15.4)$ pb,
respectively, where the first uncertainties are statistical and the
second systematic.

\end{abstract}


\maketitle

A charged charmoniumlike structure, $Z_c(3900)$, was observed in the
$\pi^\pm\jpsi$ final states by the BESIII and Belle experiments~\cite{zc3900, zc3900belle}.
Subsequently, several analogous structures were
reported and confirmed by different experiments~\cite{zc4020,zc4025,zc3885,zc4200belle,zc4430lhcb}.
These
observations inspired extensive discussions of their nature,
and the reasonable interprestations are tetraquark states,
molecular or hadroquarkonium states~\cite{tetraquark,ZcNature1, ZcNature2, ZcNature3, ZcNature4, ZcNature5, ZcNature6},
due to these structures carrying charge and
prominently decaying into a pion and a conventional charmonium state.
More recently, the neutral partners of these charmoniumlike structures
were observed~\cite{zc4020n,zc3900n,zc4025n,zc3885n}, which indicate
the isotriplet property of these structures and hint of a new hadron spectroscopy.

By replacing the $c\bar{c}$ pair in the $Z_c$ structure with an $s\bar{s}$, it is possible
to consider an analogous $Z_s$ structure.
Similar to $Y(4260)\rightarrow J/\psi\pi^+\pi^-$ in which
the $Z_c(3900)$ was observed~\cite{zc3900, zc3900belle},
the process $\phi(2170)\ra\phi\pip\pin$ is considered as a unique place to search for the $Z_s$ structure,
as the $\phi(2170)$ is regarded as the strangeoniumlike states analogy to
$Y(4260)$ in charmonium sector~\cite{y2170}.
Furthermore, the conventional isosinglet $s\bar{s}$ state decaying
into $\phi\pi$ is suppressed by the conservation of isospin symmetry, while
for a conventional meson composed of $u$, $d$ quarks, the $\phi\pi$
decay mode is strongly suppressed by the Okubo-Zweig-Iizuka (OZI) rule~\cite{ozi}.
Therefore, it is of interest to perform an experimental search for the
strangeoniumlike structure $Z_s$ since its observation may imply the
existence of an exotic state.

In this article, we present a search for the $Z_s$
structure in the process $e^+e^-\rightarrow \phi\pi\pi$
using a data sample corresponding to an integrated luminosity of
$(108.49\pm0.75)$ pb$^{-1}$~\cite{luminosity}, taken at a
center-of-mass energy of 2.125 GeV with the BESIII detector.
Since the observed $Z_c(3900)$~\cite{zc3900,zc3900belle} and $Z_c(3885)$~\cite{zc3885} are close to the
$D^*\bar{D}$ mass threshold and have a narrow width,
the search for a narrow width $Z_s$ structure around
the $K^*\bar{K}$ mass threshold ($1.4$ GeV$/c^{2}$) in the $\phi\pi$ mass spectrum
allows us to test the novel scenario of
the initial single pion emission mechanism (ISPE)~\cite{xliu}.

The BESIII detector~\cite{bes3} is a magnetic spectrometer located at
the Beijing Electron Position Collider (BEPCII), which is a
double-ring $e^+e^-$ collider with a peak luminosity of
$10^{33} ~\rm{cm}^{-2}\rm{s}^{-1}$ at a center-of-mass energy
of $3.773$ GeV. The cylindrical core of the BESIII detector consists of a
helium-based multilayer drift chamber (MDC), a plastic scintillator
time-of-flight system (TOF), and a CsI(Tl) electromagnetic calorimeter
(EMC), which are all immersed in a superconducting solenoidal magnet
providing a 1.0~T magnetic field. The solenoid is
supported by an octagonal flux-return yoke with resistive plate
counter muon identifier (MUC) modules interleaved with steel. The acceptance
of charged particles is 93\% over 4$\pi$ solid angle. The
charged-particle momentum resolution at $1$ GeV/$c$ is $0.5\%$, and the
specific energy loss ($dE/dx$) resolution is $6\%$.
The EMC measures photon energies with a
resolution of $2.5\%$ $(5\%)$ at $1$ GeV in the barrel (end caps) region. The time
resolution of TOF is 80~ps in the barrel and 110~ps in the end caps.
The position resolution in the MUC is better than 2 cm.

The {\sc GEANT}4-based~\cite{geant4} Monte Carlo (MC) simulation software, which
includes the geometric description of the BESIII detector and the
detector response, is used to determine the detection efficiencies and
estimate backgrounds.
To simulate the $e^+e^-\ra\phi\pi\pi$ process,
the lineshape reported by BaBar~\cite{xsecbabar} is adopted.
Intermediate states in the simulation of $e^+e^-\ra\phi\pi\pi$ process are modeled
according to the BESIII data as described later.

Candidate events of $e^+e^-\rightarrow \phi\pi^{+}\pi^{-}$ ($\phi\ar K^+ K^-$)
are required to have three or four charged
tracks. Charged tracks are reconstructed from hits in the MDC within
the polar angle range $|\cos\theta|<0.93$.  The tracks are required to
pass the interaction point within 10 cm along the beam direction and
within 1 cm in the plane perpendicular to the beam. For each charged
track, the TOF and the $dE/dx$ information are combined to
form particle identification (PID) confidence levels (C.L.) for the $\pi$,
$K$, and $p$ hypotheses, and the particle type with the highest
C.L. is assigned to each track.
Two pions with opposite charges
and at least one kaon are required to be identified.  A one-constraint (1$C$)
kinematic fit is performed under the hypothesis that the
$K\pi^{+}\pi^{-}$ missing mass corresponds to the kaon mass, and the
corresponding $\chi^2$, denoted as $\chi^2_{1C}(\pi^{+}\pi^{-} KK_{\rm {miss}})$, is required
to be less than 10. For events with two reconstructed and identified kaons, the
combination with the smaller $\chi^2_{1C}(\pi^{+}\pi^{-} KK_{\rm {miss}})$ is retained.

Candidate events of $e^+e^- \rightarrow \phi\pi^{0}\pi^{0}$ ($\phi\ar K^+ K^-$,
$\pi^{0} \rightarrow \gamma \gamma$) are required
to have one or two charged tracks and at least four photon candidates.
Photon candidates are reconstructed from isolated showers in the EMC,
and the corresponding energies are required to be at least 25 MeV in the
barrel ($|\cos\theta|<0.80$) or 50 MeV in the end caps
($0.86<|\cos\theta|<0.92$). To eliminate showers associated with
charged particles, the angle between the cluster and the nearest
charged track must be larger than 10 degrees.  An EMC cluster timing
requirement of $0 \le t \le 700$ ns is also applied to suppress
electronic noise and energy deposits unrelated to the event.  At least
one kaon is required to be identified. A 1$C$ kinematic fit is then
performed under the hypothesis that the $K 4\gamma$ missing mass is
the kaon mass.
For events with two identified kaons or more than four photons, the combination
with the smallest $\chi^2_{1C}(4 \gamma KK_{\rm miss})$ is retained
and required to be less than 20. The four selected photons are grouped into
pairs to form $\pi^0$ mesons. Two $\pi^0$ candidates are then selected
by minimizing the quantity
$(M(\gamma\gamma)_1-m_{\pi^{0}})^{2}+(M(\gamma\gamma)_2-m_{\pi^{0}})^{2}$,
where $m_{\pi^{0}}$ is the nominal $\pi^{0}$ mass from Particle Data Group (PDG)~\cite{pdg}.  In
order to select a clean sample, both $M(\gamma\gamma)_1$ and
$M(\gamma\gamma)_2$ are required to be within $\pm20$ MeV/$c^2$ of $m_{\pi^{0}}$.

\begin{figure}[!htbp]
\centering
 \includegraphics[width=0.25\textwidth]{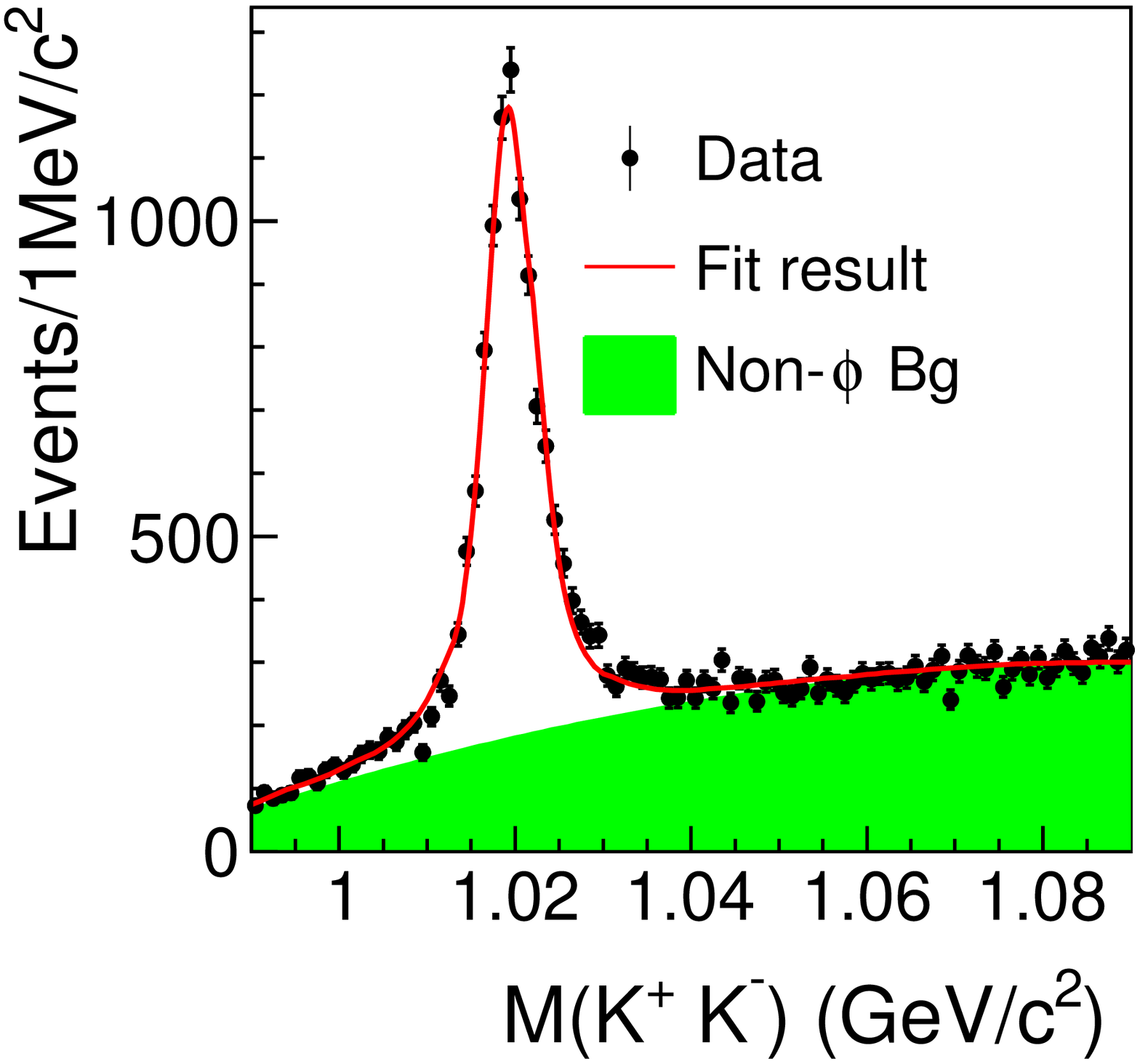}
\put(-35,95){\bf (a)}
 \includegraphics[width=0.25\textwidth]{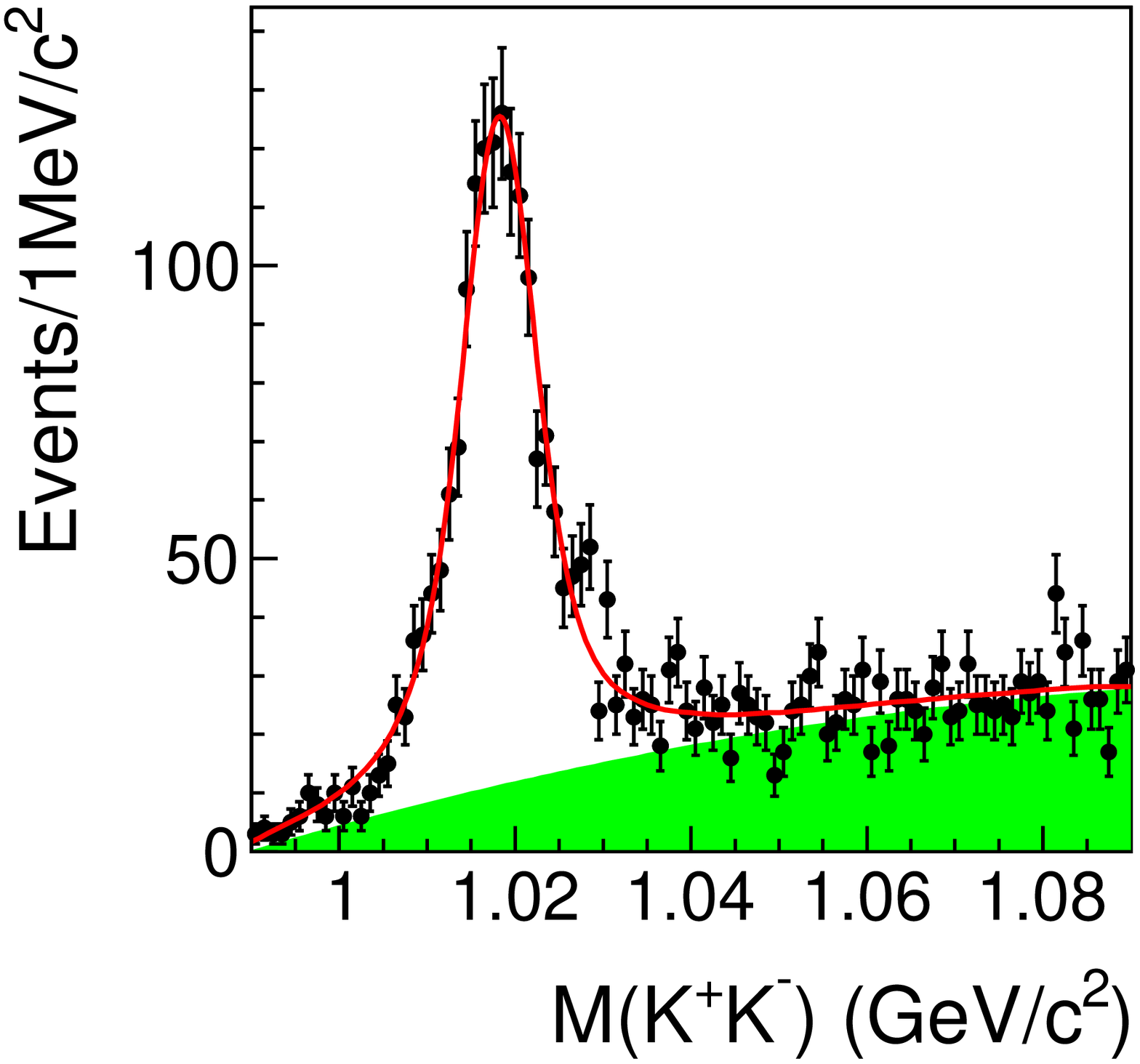}
\put(-35,95){\bf (b)}
\caption{\label{fig:Mkk_fit} Invariant mass distributions of $K^+K^-$
for (a) $e^+e^-\rightarrow K^+K^-\pi^{+}\pi^{-}$ and (b)
$e^+e^-\rightarrow K^+K^-\pi^{0}\pi^{0}$ events. The dots with error bars
are data, the solid lines are the fit results
and the shaded parts are the combinatorial backgrounds obtained from fits.}
\end{figure}

\begin{figure}[!htbp]
\centering
 \includegraphics[width=0.25\textwidth]{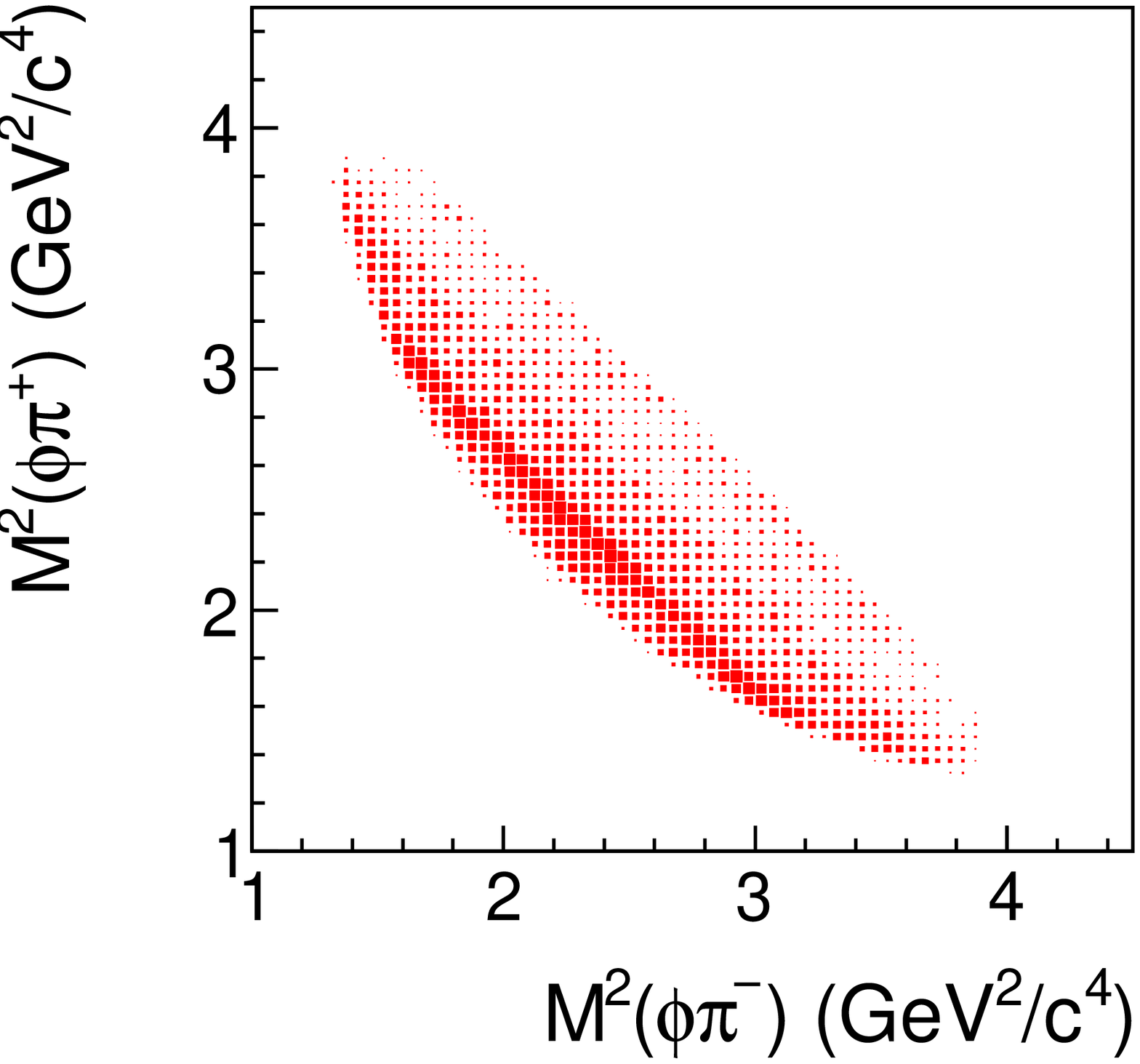}
\put(-35,95){\bf (a)}
 \includegraphics[width=0.25\textwidth]{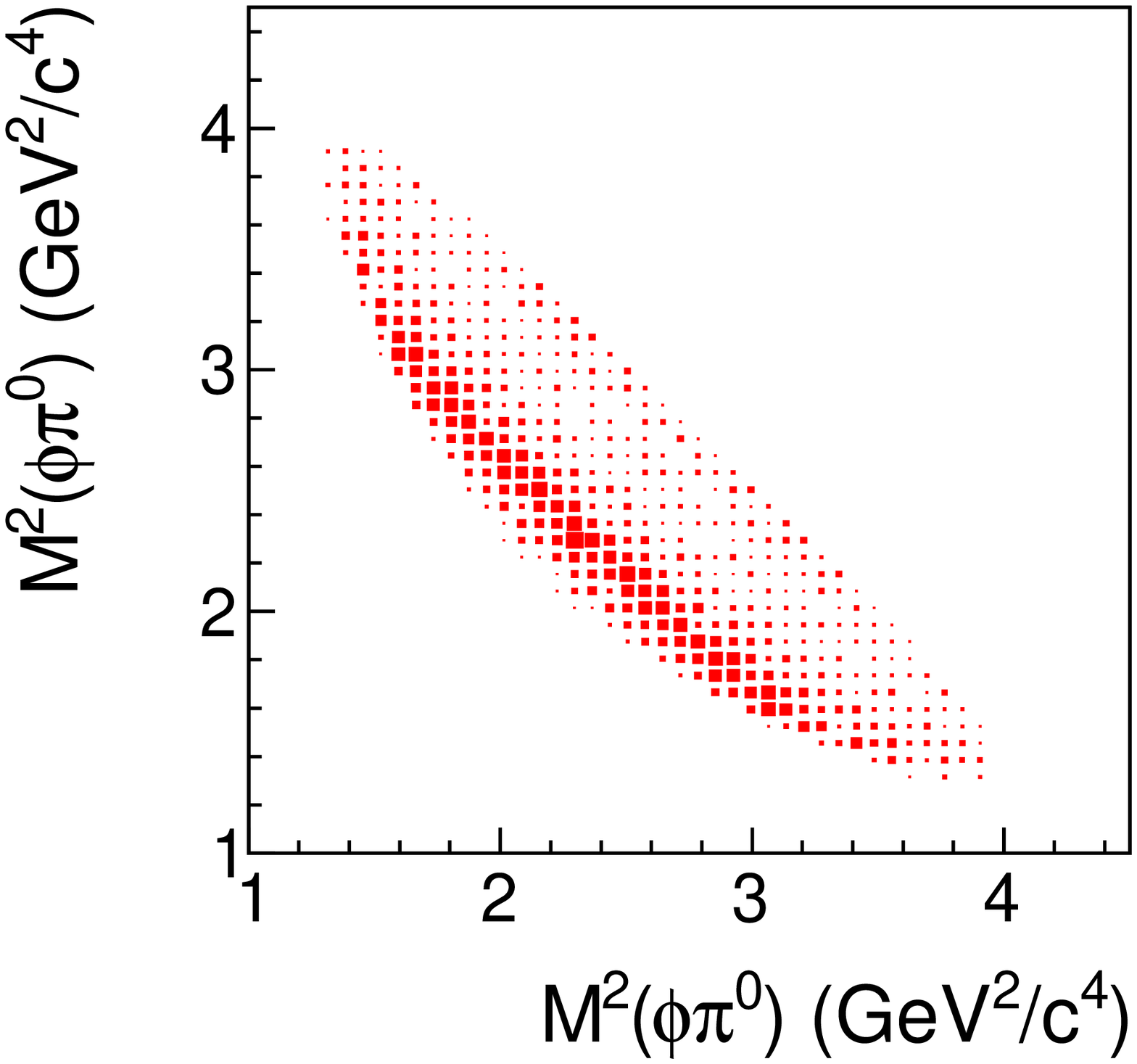}
\put(-35,95){\bf (b)}

 \includegraphics[width=0.25\textwidth]{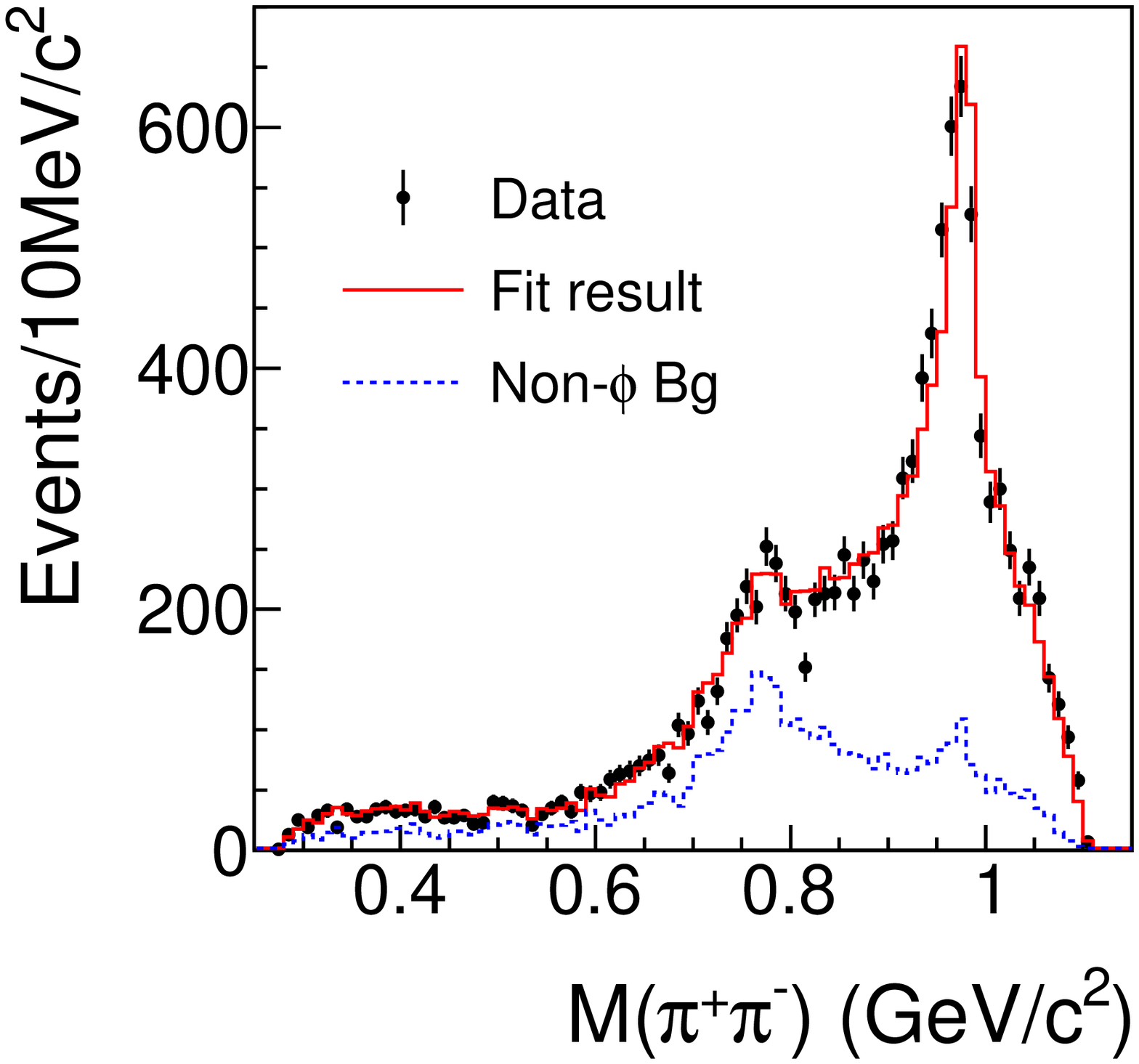}
\put(-55,95){\bf (c)}
 \includegraphics[width=0.25\textwidth]{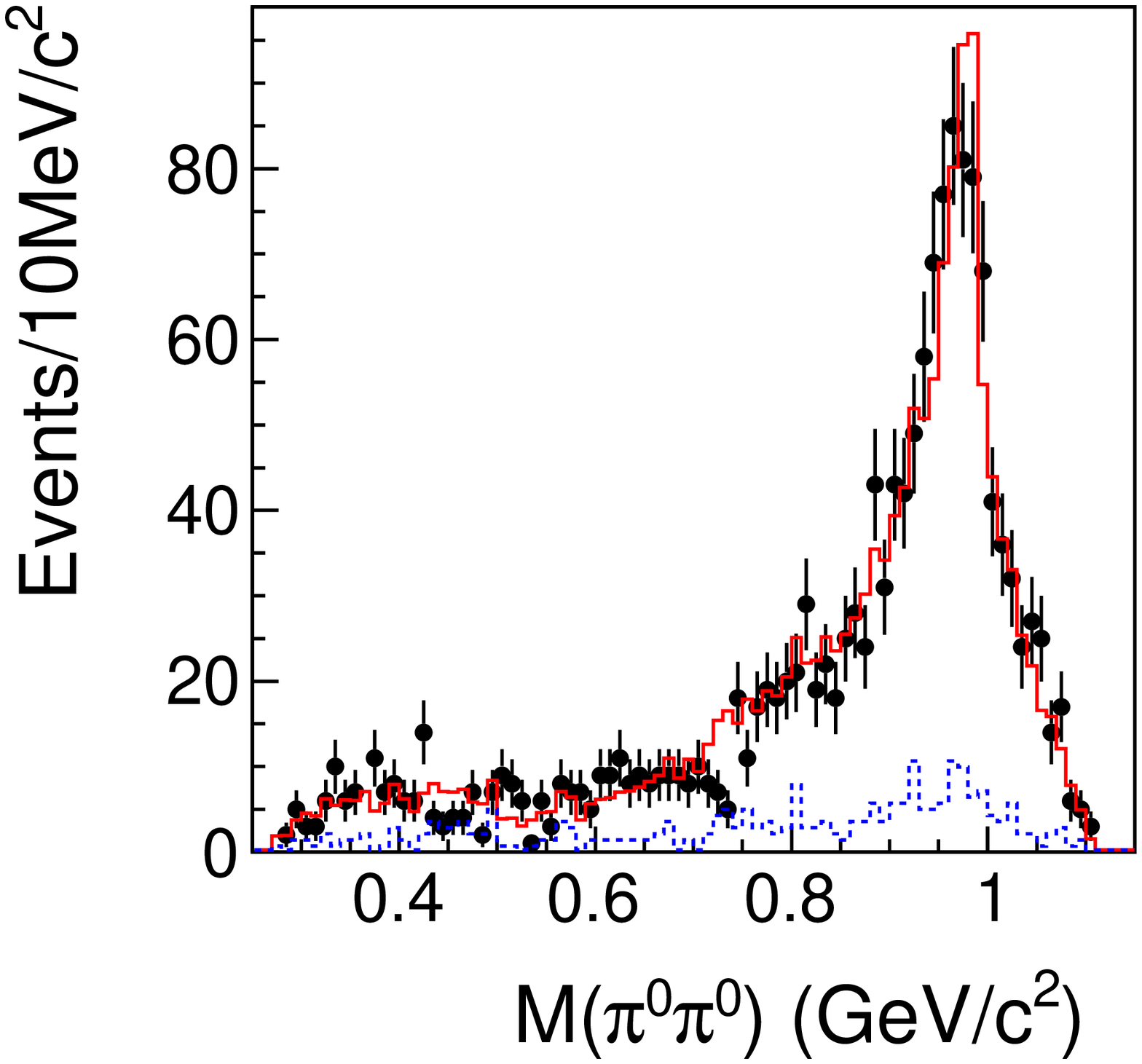}
\put(-55,95){\bf (d)}
 \caption{\label{fig:dalitz}
	Dalitz plots for
	(a) $e^+e^-\ar\phi\pi^{+}\pi^{-}$ and (b) $e^+e^-\ar\phi\pi^{0}\pi^{0}$
	candidate events and invariant mass distributions of (c)
	$\pi^{+}\pi^{-}$ and (d) $\pi^{0}\pi^{0}$. The dots with error bars
	are data, the dotted histograms are non-$\phi$ backgrounds estimated from
	$\phi$ sidebands, and the solid histograms are the sum of the projections
	of the amplitude analysis results and non-$\phi$ backgrounds.
	Each $e^+e^-\ra\phi\pi^0\pi^0$ event contributes two entries for (b).
 }
\end{figure}

After applying the above selection criteria, the
$K^+K^-$ invariant mass, $M(K^+K^-)$, is computed using the
four-momenta of the reconstructed $K$ and $K_\text{miss}$ from the
kinematic fit.
The $M(K^+K^-)$ spectra for the selected candidate events are shown in Figs.~\ref{fig:Mkk_fit}(a)
and \ref{fig:Mkk_fit}(b), where $\phi$ signals are clearly seen. The Dalitz plots of
the $\phi\pip\pin$ and $\phi\pi^{0}\pi^{0}$ events are shown in
Figs.~\ref{fig:dalitz}(a) and \ref{fig:dalitz}(b), respectively, where the
$M(K^+K^-)$ is required to be in the $\phi$ mass range,
$|M(K^+K^-)-m_{\phi}|<0.01$ GeV/$c^2$, and $m_{\phi}$ is the nominal
$\phi$ mass from PDG~\cite{pdg}. The apparent
structures are from the decay processes $e^+e^-\rightarrow \phi f_0(980)$
with $f_0(980)$ decaying to $\pi^{+}\pi^{-}$ or $\pi^{0}\pi^{0}$ final states,
which are also clearly indicated in the $\pi\pi$
invariant mass spectra, $M(\pi\pi)$, displayed in Figs.~\ref{fig:dalitz}(c) and \ref{fig:dalitz}(d).
There is a clear structure around $\rho$ mass region in the $\pi\pi$ mass
spectrum in the $K^+K^-\pi^{+}\pi^{-}$ channel. In addition,
$K^*(892)K^{\mp}\pi^{\pm}$ events also contaminate the charged process.
The contributions from those non-$\phi$ backgrounds are described by
the events in the $\phi$ sideband regions, $0.995<M(K^+K^-)<1.005$ and
$1.035<M(K^+K^-)<1.045$ GeV/$c^2$, and are normalized according to the fitted
intensities in Fig.~\ref{fig:Mkk_fit}.  The $M(\pi\pi)$ distributions
of $\phi$ sideband events are represented by the
dotted lines in Figs.~\ref{fig:dalitz}(c) and (d).

\begin{figure}[!htbp]
\centering
    \includegraphics[width=0.25\textwidth]{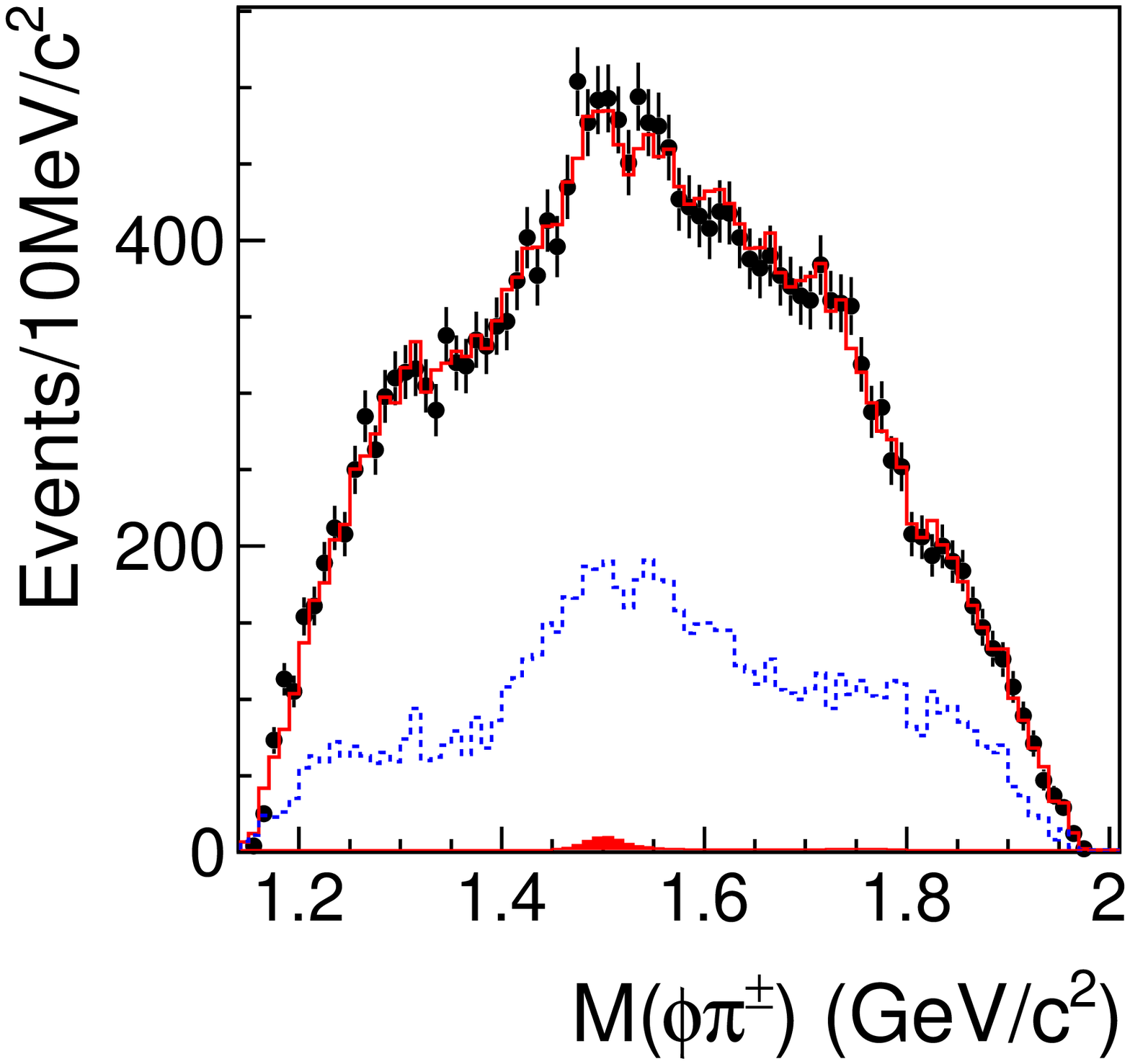}
\put(-95,98){\bf (a)}
    \includegraphics[width=0.25\textwidth]{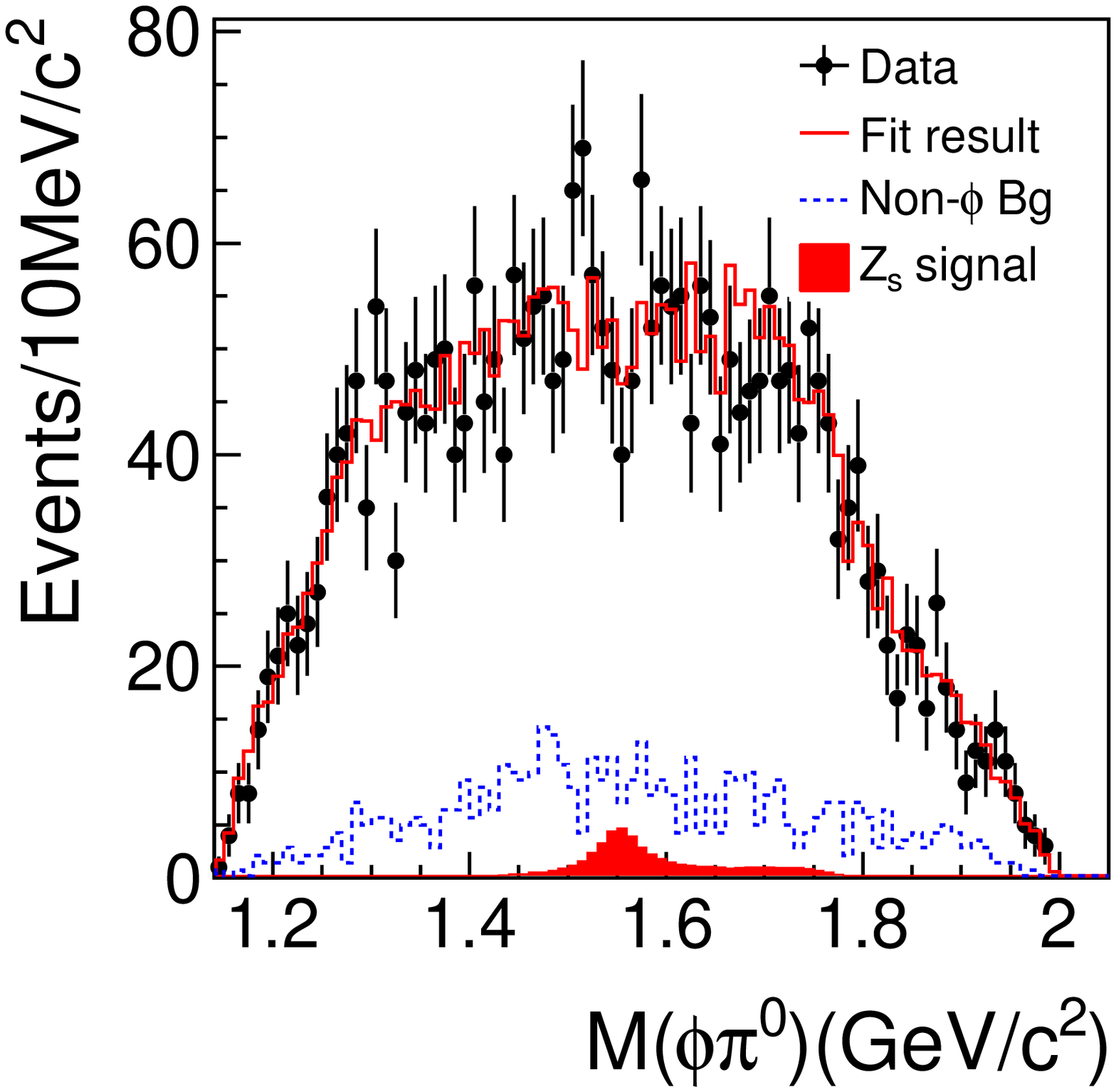}
\put(-95,98){\bf (b)}
\caption{\label{fig:mphipi} Invariant mass distributions of (a)
  $M(\phi\pi^{\pm})$ and (b) $M(\phi\pi^{0})$ for
  $\phi\pi\pi$ candidate events. The dots with error bars are data,
the solid histograms are the projections of the amplitude analysis results
	including the contributions from $Z_s\ra\phi\pi$ process with the mass and width of
	$Z^{\pm}_s$ ($Z_s^0$) assumed to be 1.5 (1.55) GeV/$c^2$ and 50 MeV for the case of $J^P=1^+$,
	the dashed histograms are non-$\phi$ backgrounds,
  and the shaded histograms are the $Z_s$ signal.}
\end{figure}

The mass spectra of the $\phi$ candidate paired with $\pi$
are shown in Fig.~\ref{fig:mphipi}. 
There is no evidence of structures in the entire $\phi\pi$ region.
To describe the $M(\pi\pi)$ spectrum,
an amplitude analysis
on $e^+e^-\ra\phi\pi\pi$ is performed using the relativistic convariant tensor
amplitude method~\cite{pwa}.

\begin{figure*}[!hbtp]
\centering
 \includegraphics[width=1\textwidth]{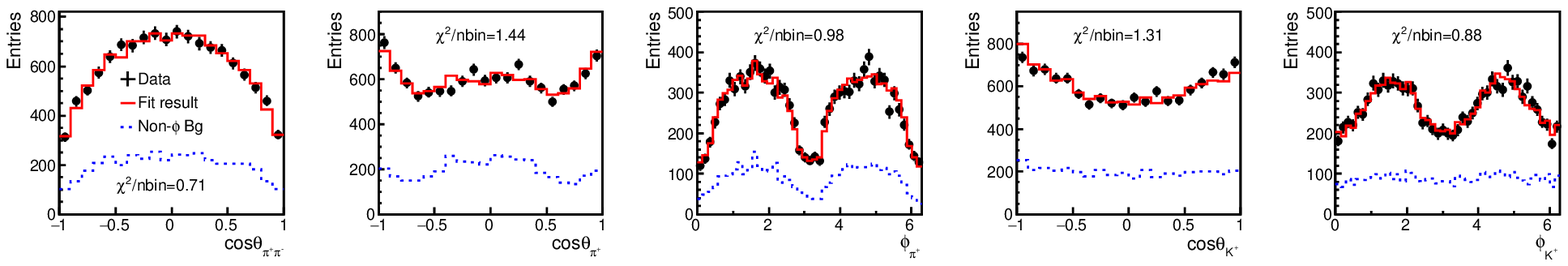}
\put(-425,75){(a)}
\put(-330,75){(b)}
\put(-222,75){(c)}
\put(-122,75){(d)}
\put(-18 ,75){(e)}

 \includegraphics[width=1\textwidth]{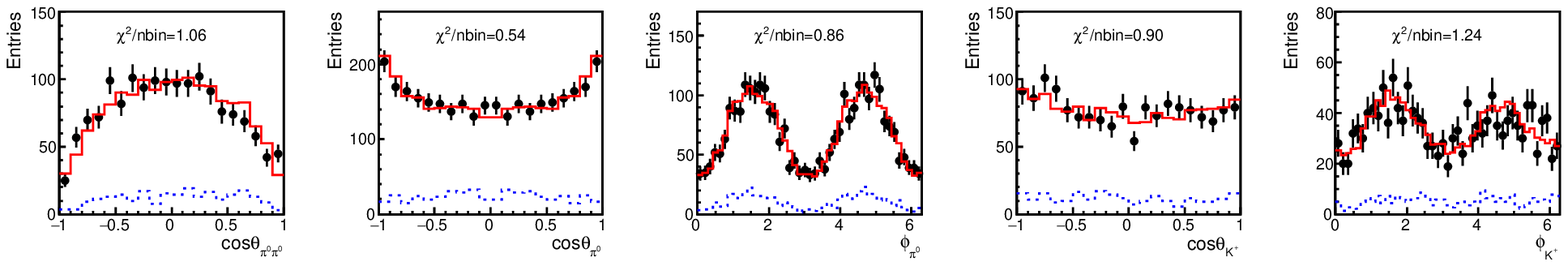}
\put(-425,75){(f)}
\put(-330,75){(g)}
\put(-222,75){(h)}
\put(-122,75){(i)}
\put(-18 ,75){(j)}
\caption{\label{fig:angular}
Angular distributions for $e^+e^-\ar\phi\pi^{+}\pi^{-}$ (a-e) and $e^+e^-\ar\phi\pi^{0}\pi^{0}$ (f-j).
For (g) and (h), there are two entries for each event due to the two identical $\pio$s in $e^+e^-\ra\phi\pio\pio$ process.
The dots with error bars
are data, the dotted histograms are non-$\phi$ backgrounds estimated from $\phi$ sidebands,
and the solid histograms are the sum of the backgrounds and the fit projections.
$\cos\theta_{\pi\pi}$ is the polar
angle of $\pi\pi$ in the rest frame of $e^+e^-$ annihilation, $\cos\theta_{\pi}$ 
($\cos\theta_{K}$) and $\phi_\pi$ ($\phi_K$) are the polar angle and azimuthal
angle of $\pi$ $(K^{+})$ in the $\pi\pi$ ($K^+K^-$) system. 
}
\end{figure*}

The $e^+e^-\ra\phi\pi\pi$ process can be described by four subprocesses:
$e^+e^-\ra\phi\sigma$, $\phi f_0(980)$, $\phi f_0(1370)$, and $\phi f_2(1270)$.
$\sigma$ is described with the form used fitting $\pi\pi$ elastic scattering data~\cite{refsigma},
$f_0(980)$ is described with a Flatt\'e formula~\cite{flatte}, and others are described
with relativistic Breit-Winger (BW) function.
The resonance parameters are fixed on
the values determined in previous BES results~\cite{pwaformula1, pwaformula2}.
Non-$\phi$ backgrounds estimated from the $\phi$ sidebands are represented by a
non-interfering term.
The projections of nominal amplitude analysis
results on the $M(\pi\pi)$
distributions are shown as the solid lines in
Figs.~\ref{fig:dalitz}(c) and \ref{fig:dalitz}(d).
The comparisons of angular distributions between data and the amplitude
analysis projections for these two interested processes are also
displayed in Fig.~\ref{fig:angular}. To illustrate the fit quality,
we present a $\chi^2$ test for each distribution ($\chi^2/nbin$), where $nbin$ is the number of bins.
In general the values of $\chi^2/nbin$ are around 1, which indicates that the amplitude
analysis results provide a reasonable description of data.

To estimate the statistical significance for each component,
alternative fits by excluding the corresponding amplitude are performed.
The statistical significance is then determined by the changes of the
log likelihood values and the number of degrees of freedom.
The statistical significances of all these states are found to be larger than 5$\sigma$.
A full partial wave analysis of $e^+e^-\ra K^+K^-\pi^+\pi^-$ is in progress with
more statistics taken at different energy points around $Y(2175)$
at BESIII, in which detailed results will be presented.

With a hypothesis of $J^{P}=1^+$, the contribution of $Z_s$ is examined by introducing
an additional component in the amplitude analysis. To simplify the analysis, we neglect
the D-wave and assume that the contribution is only from the S-wave amplitude.
The $Z_s$ is parameterized as
a relativistic BW function in the $\phi\pi$ system. As the mass and width of the state
are unknown, we have tested signals with masses of 1.2-1.95 GeV/$c^2$ in steps of 0.05 GeV/$c^2$.
For the width, values of 10, 20, and 50 MeV are combined with each mass.
With these different signal hypotheses, we performed the fit to data and found,
in general, that the observed statistical significances are less than 3$\sigma$ in the explored region.
For $e^+e^-\ra\phi\pi^+\pi^-$, the maximum local significance is $2.7\sigma$ in the case of 
$M(Z^{\pm}_s)=1.5$ GeV/$c^2$ and $\Gamma(Z^{\pm}_s)=50$ MeV,
which becomes to be $2.1\sigma$ after taking the systematic uncertainty into account,
and the signal yields are
determined to be 46.9$\pm$21.6.
While for $e^+e^-\ra\phi\pi^0\pi^0$,
the maximum local significance is $3.3\sigma$ in the case of 
$M(Z^0_s)=1.55$ GeV/$c^2$ and $\Gamma(Z^0_s)=50$ MeV,
which becomes to be $2.8\sigma$ after taking the systematic uncertainty into account,
and the signal yields are
determined to be 25.2$\pm$8.9.
The corresponding projections of the amplitude analysis results on $M(\phi\pi^\pm)$ and
$M(\phi\pi^0)$ are shown in Figs.~\ref{fig:mphipi}(a) and \ref{fig:mphipi}(b), respectively.

In the determination of the upper limits on the number of $Z_s$
($N^{UL}$) for different scenarios,
the same approach as that in Ref.~\cite{ulmethod} is used. For each case, the statistical
uncertainty is used to determine the 90\% C.L. deviation, and added to the
nominal yields to obtain the corresponding upper limit on the number of $Z_s$ signals.

\begin{figure}[!htbp]
\centering
\includegraphics[width=0.5\textwidth]{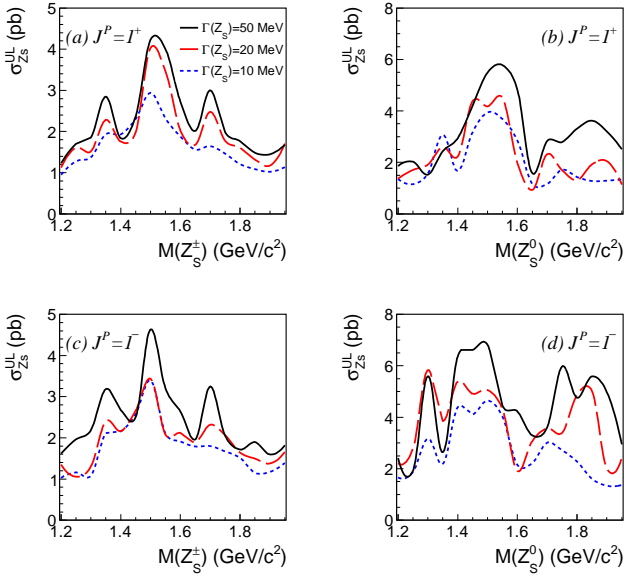}
\caption{\label{fig:differentw-zs} The upper limits at 90\% C.L. on
the differential cross sections of $Z_s$ as a function of assumed signal peak
mass for the cases (a) $J^P=1^+$ of $Z_s^{\pm}$, (b) $J^P=1^+$ of $Z_s^{0}$,
(c) $J^P=1^-$ of $Z_s^{\pm}$, and (d) $J^P=1^-$ of $Z_s^{0}$.
The dotted, dashed and solid lines are the results of $\Gamma$=10, 20, and 50 MeV, respectively.
}
\end{figure}

The systematic uncertainties on the upper limit of $Z_s$ signal yields associated
with $\phi$ sideband range and the nominal $\phi\pi\pi$ model,
estimated by varying the resonance parameters or
replacing the $f_0(1370)$ component with a phase space process,
are considered by performing
alternative fits and taking the maximum value of $N^{UL}$ as the upper limit,
while the other systematic uncertainties are taken into account by
dividing the factor $(1-\delta_{syst.})$, where
$\delta_{syst.}$ is total systematic uncertainties, described in
detail later. With the detection efficiency obtained from the
dedicated MC simulation for each $Z_s$ hypothesis, 
the upper limit on the cross section  is calculated with
\begin{equation}
 \sigma^{UL}_{{Z_s}}(e^{+}e^{-} \rightarrow Z_{s} \pi, Z_{s} \rightarrow
 \phi\pi)=\frac{N^{UL}}{\mathcal{L}(1+\delta)(1-\delta_{syst.})\varepsilon
{\mathcal B}},
\label{equ:born-cross}
\end{equation}
where $\mathcal{L}$ is the integrated luminosity of the data taken at
2.125 GeV, and determined to be $(108.49\pm0.75)$ pb$^{-1}$~\cite{luminosity} from
large-angle Bhabha scattering events;
$(1+\delta)$ is
a radiative correction factor calculated to the second-order in
QED~\cite{radcorrect} by assuming that the line shape follows the
measured cross section of the BaBar experiment~\cite{xsecbabar},
determined as 0.982 and 0.986 for the $e^+e^-\ra\phi\pip\pin$ and
$\phi\pio\pio$ channels, respectively;
$\varepsilon$ is the detection efficiency;
and ${\mathcal B}$ is either ${\mathcal B}(\phi \to K^+K^-)$ for
$\phi\pip\pin$ or ${\mathcal B}(\phi \to K^+K^-)\times{\mathcal
B}^2(\pi^0 \to \gamma \gamma)$ for $\phi\pio\pio$~\cite{pdg}.
The corresponding upper limits on the differential cross sections of $Z_{s}$ production
as a function of the assumed mass of $Z_s$ with
different width scenario are shown in Figs.~\ref{fig:differentw-zs}(a) and \ref{fig:differentw-zs}(b).

In addition, we performed the alternative amplitude analysis by assuming $J^P=1^-$
to explore the $Z_s$ contribution to the data. With the same approach as described above,
the upper limits on the differential cross sections of $Z_{s}$ production
as a function of the assumed mass of $Z_s$ with different width scenario are also estimated at
90\% C.L., which are displayed in Figs.~\ref{fig:differentw-zs}(c) and \ref{fig:differentw-zs}(d).

The $e^+e^-\rightarrow\phi\pi\pi$ signal yields are obtained from
extended unbinned maximum likelihood fits to the $M(K^+K^-)$ distributions.
In the fit, the $\phi$ peak is modeled as the signal MC simulated
shape convoluted with a Gaussian function to account for the mass
resolution difference between data and MC simulation, while the
background is described by a second-order polynomial function.  The
fits to $M(K^+K^-)$ spectra, shown in Figs.~\ref{fig:Mkk_fit}(a) and \ref{fig:Mkk_fit}(b),
yield $(9421\pm138)$ $\phi\pi^{+}\pi^{-}$ and $(1649\pm60)$
$\phi\pi^{0}\pi^{0}$ events. The detection efficiencies are
($41.2\pm0.1$)\% and ($13.7\pm0.1$)\%, respectively, obtained from the
signal MC samples generated according to the nominal amplitude analysis results.
The cross sections
for $e^+e^-\rightarrow\phi \pi^{+}\pi^{-}$ and
$e^+e^-\rightarrow\phi\pi^{0}\pi^{0}$ are determined to be
$(436.2\pm6.4)$ pb and $(237.0\pm8.6)$ pb, respectively.

Sources of systematic uncertainties and their corresponding
contributions to the measurements of the cross sections are summarized
in Table~\ref{tab:syserr}. The uncertainties of the MDC tracking
efficiency for each charged kaon and pion and the photon selection
efficiency are studied with a control sample $e^{+}e^{-}\rightarrow K^+K^-\pi^{+}
\pi^{-}$ taken at the energy of $2.125$ GeV and a
control sample of $e^{+}e^{-}\rightarrow \pi^{+}\pi^{-}\pi^{0}$ taken at the energy of $3.097$ GeV,
respectively, and the differences between data and MC simulation are
less than $1.5\%$ per charged track and $1.0\%$ per photon.
Similarly, the uncertainties related to the pion and kaon PID
efficiencies are also studied with the sample $e^+e^-\ar K^+K^-\pi^{+}\pi^{-}$,
and the average differences of the PID
efficiencies between data and MC simulation are determined to be 3\% and
1\% for each charged kaon and pion, respectively, which are taken as
the systematic uncertainties.

Uncertainties associated with kinematic fits come from the
inconsistency of the track helix parameters between data and MC simulation.  The helix
parameters for the charged tracks of MC samples
are corrected to eliminate the inconsistency, as described in
Ref.~\cite{syskifit}, and the agreement of $\chi^2$ distributions
between data and MC simulation is much improved. We take
half of the differences on the selection efficiencies with and without
the correction as the systematic uncertainties, which are 2.1\% for
$\phi\pi^{+}\pi^{-}$ and 0.1\% for $\phi\pi^{0}\pi^{0}$ channels, respectively.
The difference of the
selection efficiencies associated with the $\pi^{0}$ mass window requirement
between data and MC simulation is estimated to be about
$0.1\%$, which is taken as the systematic uncertainty for the mode
$e^{+}e^{-}\rightarrow \phi \pi^{0} \pi^{0}$.
The systematic uncertainty on the $Z_s$ production associated with the $M(K^+K^-)$
mass window is estimated by alternative fits varing the cut by
1$\sigma$ and found to be 1.5\%.

In the measurement of the cross section for $e^+e^-\ra\phi\pi\pi$, the
nominal fit range for $M(K^+K^-)$ is (0.99, 1.09) GeV/$c^2$.
Alternative fits are performed by varying the fitting range.  The
maximum changes on the calculated cross sections are assigned as the
uncertainties from the fitting range.
The uncertainties associated with
the background shape in the fits to $M(K^+K^-)$ are estimated with alternative fits
by changing the second-order polynomial function to a third-order Chebychev polynomial function.
Alternative fits to $M(K^+K^-)$ are performed by removing the smeared resolution function to estimate
the uncertainties associated with the $\phi$ signal shape.
The resultant differences are assigned as the systematic uncertainties.
In the amplitude analysis, alternative fits are performed by
varying the parameters of resonances according to the previous BES
results~\cite{pwaformula1, pwaformula2} or replacing the component
of $f_0(1370)$ intermediate state with a phase space process with
$J^{PC} = 0^{++}$. The model with the maximum changes on the
log-likelihood values are used to estimated the systematic
uncertainties associated with the model.

The branching fractions of the
intermediate processes $\phi\rightarrow K^+K^-$ [$(49.2\pm0.5)$\%] and
$\pi^{0}\ar\gamma\gamma$ [$(98.823\pm0.034)$\%] are taken from the
PDG~\cite{pdg}, where the overall uncertainty, $1.0\%$,
is taken as the systematic uncertainty.
The luminosity is determined to be $(108.49\pm0.75)$ pb$^{-1}$ in
Ref.~\cite{luminosity} with an uncertainty of $0.7\%$.
Uncertainties in the $Y(2125)$ resonance parameters and
possible distortions of the $Y(2125)$ line shape introduce small systematic
uncertainties in the radiative correction factor and the efficiency.
This is estimated using the different line shapes measured by BaBar and Belle, and the
difference in $(1+\delta)\cdot\varepsilon$ are taken as a systematic error,
1.0\% for $e^+e^-\ra\phi\pip\pin$ and 0.7\% for $e^+e^-\ra\phi\pio\pio$, respectively.

\begin{table}[!htbp]
\centering
\caption{\label{tab:syserr} Systematic uncertainties (in \%) for
		the measurements of the upper limits (uncorrelated ones)
		and cross sections. Assuming the uncertainties
		are uncorrelated, the total uncertainty is the quadratic
		sum of the individual values.}
\begin{tabular}{lcccc}\hline\hline
	Source  & $Z_s^{\pm}$ & $\phi\pi^{+}\pi^{-}$ & $Z_s^0$ & $\phi\pi^{0}\pi^{0}$	\\\hline
	MDC tracking          &    4.5   &    4.5   &  1.5  &  1.5   \\
	Photon detection      &    ...   &    ...   &   4   &   4    \\
	K PID                 &     3    &     3    &   3   &   3    \\
	$\pi$ PID             &     2    &     2    &  ...  &  ...   \\
	Kinematic fit 			  &    2.1   &    2.1   &  0.1  &  0.1   \\
	$\pi^{0}$ mass window &    ...   &    ...   &  0.1  &  0.1   \\
	$K^+K^-$ mass window  &    1.5   &    ...   &  1.5  &  ...   \\
	Fitting range 			  &    ...   &    0.1   &  ...  &  1.4   \\
	Signal shape 		  		&    ...   &    1.5   &  ...  &  2.3   \\
	Background shape 		  &    ...   &    1.3   &  ...  &  2.0   \\
	Model uncertainty     &    ...   &    0.8   &  ...  &  1.3   \\
	Branching fractions   &    1.0   &    1.0   &  1.0  &  1.0   \\
	Integrated luminosity &    0.7   &    0.7   &  0.7  &  0.7   \\
	ISR									  &    1.0   &    1.0   &  0.7  &  0.7   \\\hline
	Total 			          &    6.5   &    6.9   &  5.6  &  6.5	 \\\hline\hline
  \end{tabular}
\end{table}

In summary, a search for a strangeoniumlike structure, $Z_s$, in the process $e^+e^-
\rightarrow \phi\pi\pi$ is performed using $108$ pb$^{-1}$ of data
collected with the BESIII detector at 2.125 GeV.
No $Z_{s}$ signal is observed in the $\phi\pi$ invariant mass spectrum,
and corresponding upper limits on the cross sections of $Z_s$ production
at the 90\% C.L. are determined 
for different mass and width hypotheses,
as displayed in Fig.~\ref{fig:differentw-zs}. The results around 1.4 GeV/$c^2$ indicate the
ISPE mechanism at $K^{*}\bar{K}$ threshold is not as significant as
predicted in Ref.~\cite{xliu}.
Further study with larger statistics is essential to examine the
existence of the $Z_{s}$ structure and test the ISPE mechanism.

In addition, the cross sections for
$e^+e^-\rightarrow \phi \pi^{+}\pi^{-}$ and
$e^+e^-\rightarrow \phi \pi^{0}\pi^{0}$ are determined
to be $(436.2\pm6.4\pm30.1)$ pb and $(237.0\pm8.6\pm15.4)$ pb, respectively.
The measured cross sections
are consistent with previous measurements
from the BaBar
($510\pm50\pm21$ pb at 2.1125 GeV for $e^+e^-\rightarrow \phi\pi^{+}\pi^{-}$
and $195\pm50\pm14$ pb at 2.100 GeV for $e^+e^-\rightarrow \phi\pi^{0}\pi^{0}$)~\cite{xsecbabar} and
Belle experiments
($480\pm60\pm42$ pb at 2.1125 GeV for $e^+e^-\rightarrow \phi\pi^{+}\pi^{-}$)~\cite{xsecbelle}
within unicertainties.
For both measurements, the statistical uncertainties are reduced significantly.

The BESIII Collaboration thanks the staff of BEPCII and the IHEP computing center for
their strong support. This work is supported in part by National Key Basic Research
Program of China under Contract No. 2015CB856700; National Natural Science Foundation
of China (NSFC) under Contracts No. 11235011, No. 11335008, No. 11425524, No. 11625523,
No. 11635010, No. 11675184, and No. 11735014;
the Chinese Academy of Sciences (CAS) Large-Scale Scientific Facility Program;
Youth Science Foundation of China under Contract No. Y5118T005C;
the CAS Center for Excellence in Particle Physics (CCEPP); Joint Large-Scale
Scientific Facility Funds of the NSFC and CAS under Contracts No. U1332201,
No. U1532257, and No. U1532258; CAS under Contracts No. KJCX2-YW-N29, No. KJCX2-YW-N45,
and No. QYZDJ-SSW-SLH003; 100 Talents Program of CAS; National 1000 Talents Program of China;
INPAC and Shanghai Key Laboratory for Particle Physics and Cosmology; German Research
Foundation DFG under Contracts Nos. Collaborative Research Center CRC 1044, FOR 2359;
Istituto Nazionale di Fisica Nucleare, Italy;
Koninklijke Nederlandse Akademie van Wetenschappen (KNAW)
under Contract No. 530-4CDP03; Ministry of Development of Turkey under Contract
No. DPT2006K-120470; National Natural Science Foundation of China (NSFC) under
Contracts No. 11505034 and 11575077; National
Science and Technology fund; The Swedish Research Council; U. S. Department of
Energy under Contracts No. DE-FG02-05ER41374, No. DE-SC-0010118, No. DE-SC-0010504,
and No. DE-SC-0012069; University of Groningen (RuG) and the Helmholtzzentrum fuer
Schwerionenforschung GmbH (GSI), Darmstadt; and the WCU Program of National Research
Foundation of Korea under Contract No. R32-2008-000-10155-0.

\end{document}